\documentclass{raa} 
\usepackage{graphicx,times}
\usepackage{natbib}
\usepackage{amssymb,amsmath}
\bibpunct{(}{)}{;}{a}{}{,}

\def\nht{${\rm NH_3}$}

\def\nthp{${\rm N_2H^+}$}
\def\ntcs{$\rm \sigma_{NT}/c_{s}$}
\newcommand{\scouse}{{\sc scouse}}

\usepackage{hyperref}

\begin{document} 

   \title{Resolution-dependent Subsonic Non-thermal Line Dispersion Revealed by ALMA}

     \author{Nannan Yue\inst{1,2,3}, Di Li\inst{1,2,4}, Qizhou Zhang\inst{3}, Lei Zhu\inst{1}, Jonathan Henshaw\inst{5}, Diego Mardones\inst{6}, Zhiyuan Ren\inst{1}
   }

	 	\institute{ National Astronomical Observatories, CAS, Beijing 100012 { \it nnyue@nao.cas.cn; dili@nao.cas.cn }
		\and University of Chinese Academy of Sciences, Beijing 100049, China
		\and Harvard-Smithsonian Center for Astrophysics, 60 Garden Street, Cambridge, MA 02138, USA
		\and NAOC-UKZN Computational Astrophysics Centre, University of KwaZulu-Natal, Durban 4000, South Africa
		\and
		Max-Planck-Institut f\"ur Astronomie, K\"onigstuhl 17, D-69117 Heidelberg, Germany
		\and
		Departameto de Astronom\'{i}a, Universidad de Chile, Casilla 36-D, Santiago, Chile
	}

  \abstract
{We report here Atacama Large Millimeter/submillimeter Array (ALMA) N$_2$H$^+$ (1-0) images of the Orion Molecular Cloud 2 and 3 (OMC-2/3) with high angular resolution (3\arcsec\ or 1200 au) and high spatial dynamic range.
Combining dataset from the ALMA main array, ALMA Compact Array (ACA), the Nobeyama 45m Telescope, and the JVLA (providing temperature measurement on matching scales), we find that most of the dense gas in OMC-2/3 is subsonic (\ntcs = 0.62) with a mean line width ($\Delta\upsilon$) of 0.39 km s$^{-1}$ FWHM. 
This is markedly different from the majority of previous observations of massive star-forming regions. In contrast, line widths from the Nobeyama Telescope are transonic at 0.69 km s$^{-1}$ (\ntcs = 1.08). 
We demonstrated that the larger line widths obtained by the single-dish telescope arose from unresolved sub-structures within their respective beams. The dispersions from larger scales $\sigma_{ls}$ (as traced by the Nobeyama Telescope) can be decomposed into three components $\rm \sigma_{ls}^2 = \sigma_{ss}^2+ \sigma_{bm}^2+ \sigma_{rd}^2$, where small-scale $\sigma_{ss}$ is the line dispersion of each ALMA beam, bulk motion $\sigma_{bm}$ is dispersion between peak velocity of each ALMA beam, and $\sigma_{rd}$ is the residual dispersion. Such decomposition, though purely empirical, appears to be robust throughout our data cubes. Apparent supersonic line widths, commonly found in massive molecular clouds, are thus likely due to the effect of poor spatial resolution. 
The observed non-thermal line dispersion (sometimes referred to as `turbulence') transits from supersonic to subsonic at $\sim 0.05$ pc scales in OMC-2/3 region. Such transition could be commonly found with sufficient spatial (not just angular) resolution, even in regions with massive young clusters, such as Orion molecular clouds studied here.}


\titlerunning{Resolution-dependent Subsonic Turbulence Revealed by ALMA}
\authorrunning{Yue et al.}
\maketitle
%

%-------------------------------------------------------------------

\section{Introduction}
Turbulence is omnipresent in star-forming clouds. 
The common proxy for the measurement of astrophysical turbulence is line-of-sight non-thermal spectral line dispersion, i.e., the observed line dispersion subtracted by the contribution from thermal motion \citep{Larson1981}.
One of the key distinctions between clouds forming low-mass stars and those forming high-mass ones, is their line width. Surveys of dense cores harboring low-mass protostars found the non-thermal line dispersion to be often smaller than the sound speed \citep{Myers1983,Caselli2002,Rosolowsky2008}.
On the other hand, the massive cores exhibit the non-thermal line dispersion consistently larger than the sound speed \citep{Goldsmith1987,Caselli1995}.
For example, the turbulent line width of a survey toward 63 high-mass star-forming cores associated with water masers (\cite{Shirley2003}, with a resolution of $\sim$ 0.6 pc), is found to be highly supersonic, at least 4 times more turbulent than low-mass star-forming regions  
(\cite{Mardones1997,Gregersen1997}; with resolutions of around 0.06 pc).

The larger line widths observed in high-mass star-forming regions led to a theoretical picture of massive formation in which massive cores are supported by turbulence in quasi-equilibrium (see \citet{Tan2004}). A strong turbulent support manifested by larger line widths allows cores to assemble more materials to fuel high-mass star formation. The accretion rate scales as 
\begin{equation}
\dot{m_{*}} \sim \frac{c^3_{eff}}{G},
\end{equation}
where the effective sound speed $c^{3}_{eff}$ incorporates thermal gas pressure, magnetic pressure, and turbulence \citep{McKee2007}.

The Orion Molecular Cloud (OMC) is the closest region containing massive young stellar cluster, i.e., OB cluster \citep{Genzel1989,Peterson2008}. 
Multiple CO surveys of OMC have been published \citep{Shimajiri2011,Kong2018}. Large maps of dense gas tracers are also becoming common place (e.g.\ \cite{Tatematsu2008}). Recent interferometric observations of dense gas tracers prove to be particularly useful for studying the evolution of line width, down to thermal Jeans scale.
\cite{Friesen2017} released the data of GBT \nht\ survey of Gould Belt clouds with a beam size of 32\arcsec, and found that the velocity dispersion and gas kinetic temperature of $\rm NH_3$ increase with increasing star formation activity.
\cite{Li2013} presented observations of \nht\  to OMC2 and OMC3 by VLA and GBT with a resolution of 5\arcsec\ and reported the temperatures for a total 30 relatively quiescent cores. They found these cores have a mean temperature of 17 K and 83\% of cores are gravitational bound.
\citet{Svoboda2019} surveyed 12 of high-mass starless clump candidates with a beam of 0.8\arcsec\ from ALMA band 6 (1.3 mm). They suggested that the highest-density regions are not strongly supported against thermal gravitational fragmentation by turbulence nor magnetic fields. \citet{Hacar2018} investigated the Integral Shape Filament using the dense gas tracer \nthp\ with ALMA and IRAM 30m observations and identified 55 dense fibers with a median filament width FWHM of 0.035 pc.

The proximity of OMC and the increasing availability of spectral images allow for a systematic probe into the velocity dispersions of star forming clouds containing OB clusters. 
We follow up our previous high-resolution \nht\ studies of the OMC-2 and OMC-3 regions \citep{Li2013} with ALMA observations covering the same regions. The combined data sets thus provide both thermal and non-thermal line dispersion at a comparable spatial resolution and with $>$ 100 linear spatial dynamic range. Compared to \citet{Hacar2018}, our analysis utilizes temperature information at higher spatial resolution closer to that of ALMA, $\sim37$\% more sky area in OMC-2 and OMC-3, and less than half of the noise. The line widths are found to cross the crucial values of the sound speed as measured by our JVLA \nht\ observations \citep{Li2013} at scales probed by ALMA. 

Such analysis of the transition from supersonic non-thermal to sub-sonic non-thermal motion is only possible now due to the comprehensive coverage of spatial scales of this set of images of dense gases, which covers from $\sim$0.05 pc with Nobeyama to $\sim$0.006 pc with ALMA. More commonly used gas tracers, such as CO, suffer from depletion and/or optical depth effects, which prevent an accurate measurement of varying line widths. \nthp\ is one of the best dense gas tracer for studying line width since it has an isolated hyperfine component, allowing us to avoid line blending issues of \nht\ as well as quantifying line opacities. Furthermore, it is critical to have temperature measurements on matching spatial scales to separate the thermal component from line dispersion. 

We present our results as follows. Section 2 describes new observations and archival data sets, followed by analysis in Section 3 and discussions in Section 4. In Section 5, we present our main conclusions.

%====section: observations====
\section{Observations and Data Reduction}
The ALMA observations of OMC-2/3 (proposal ID 2013.1.00662.S) were carried out between November 2014 and August 2015 in Band 3. Four basebands were used, one of which covered the ${\rm N_2H^+}$ $J =1-0$ transition at 93.173 GHz with a $\sim$0.11 ${\rm km s^{-1}}$ velocity resolution. Both the 12-m main array and ACA were used to mosaic this region with 11 pointings towards OMC-3, 7 pointings towards OMC-2N and 7 pointings towards OMC-2S. The 12-m array observations consist of 14 execution blocks (EBs) in total, half of which on the OMC-2 region (on-source time of 175 minutes) and the other half on the OMC-3 region (on-source time of 174 minutes). The ACA observations include 13 EBs for OMC-3 region with an on-source time of 340 minutes, and three EBs for the OMC-2 region with an on-source time of 97 minutes. The baselines of the 12-m array observations range from 13.6 to 340.0 m, and for ACA observations the baselines are from 6.8 to 87.4 m.

The visibility data were manually calibrated for each EB with CASA 5.4.0. The $(u,v)$ datasets of the 12-m array and ACA were combined and  images were synthesized with the `tclean' task using the interactive mode. The threshold of 3 mJy and the iteration number of 1000 were used for CLEAN. Briggs weighting was set with robust of 0.5. The velocity of the data cube was centered at the rest frequency of 93.17340 GHz. The output image had a total of 200 channels with the width of 0.1 km s$^{-1}$ and the pixels had a size of 0.4\arcsec.
 Primary-beam correction was applied to the restored images. The representative FWHM of the synthesized beam for the images is 3.57\arcsec$\times$2.20\arcsec, P.A. = 82.40$^{\circ}$. The rms noise level is $\sim$ 8.3 mJy/beam in a 0.1 km s$^{-1}$ channel width. 
 Furthermore, in order to recover the missing flux in the ALMA data, we combine the single-dish ${\rm N_2H^+}$ (1-0) image \citep{Nakamura2019} from the Nobeyama observations with the ALMA images using ``casafeather" in the CASA package. The sdfactor used in casafeather is 0.25 in OMC-3 region and 0.5 in OMC-2N and OMC-2S. 
The resulting beam size is 3.46\arcsec$\times$2.20\arcsec, P.A. = 87.04$^{\circ}$ and the rms is $\sim$ 11 mJy/beam in a 0.11 km s$^{-1}$ channel width.

The single-dish data used in the combination
were taken with the Nobeyama 45m telescope in NRO Star Formation Legacy Project (PI; Fumitaka Nakamura)\footnote{
The data is from NRO Star Formation Legacy Project and retrieved from the JVO portal (\url{http://jvo.nao.ac.jp/portal/nobeyama/sfp.do}) operated by ADC/NAOJ. Nobeyama Radio Observatory(NRO) is a branch of the National Astronomical Observatory of Japan, National Institutes of Natural Sciences.}, with a 23.4\arcsec beam. The spectral resolution is 0.11 km s$^{-1}$. The details of this project are described by \cite{Nakamura2019}. The ${\rm N_2H^+}$ molecular line of Orion A was mapped in 2016 with the powerful new receiver FOREST.  FOREST is a newly-developed 100 GHz-band 4-beam dual-polarization receive and is well suited for conducting large-scale mapping observations.

The kinetic temperature map was derived from $\rm$ the combined Very Large Array (VLA) -Green Bank Telescope (GBT) images of the NH$_3$ (J, K ) = (1,1) and (2,2) inversion transitions \citep{Li2013}. 
 The combined data have a  channel width of 0.62 km s$^{-1}$ and a synthesized beam size of 5\arcsec\ FWHM. Adopted from \cite{Li2013}, the temperature maps used in this work were derived though an analytical recipe based on two line ratios, namely, the intensity ratio between the main and satellite groups within the (1,1) transition and the ratio between the integrated intensities of the (1,1) and (2,2) lines.
 Unlike most other such pipelines, this method requires no hyperfine fitting and naturally produces statistical uncertainties of the fitted temperatures, with dependence on spectral noise. In this data set, the representative 1$\sigma$ uncertainty of the resulting T$_k$ is about 1 to 2 K. The detailed procedure was described in \cite{Li2013}.

%====section: results======
\section{Analysis}
\begin{figure}
       \includegraphics[trim=0 0cm 0 0, clip, angle=0,width=15cm]{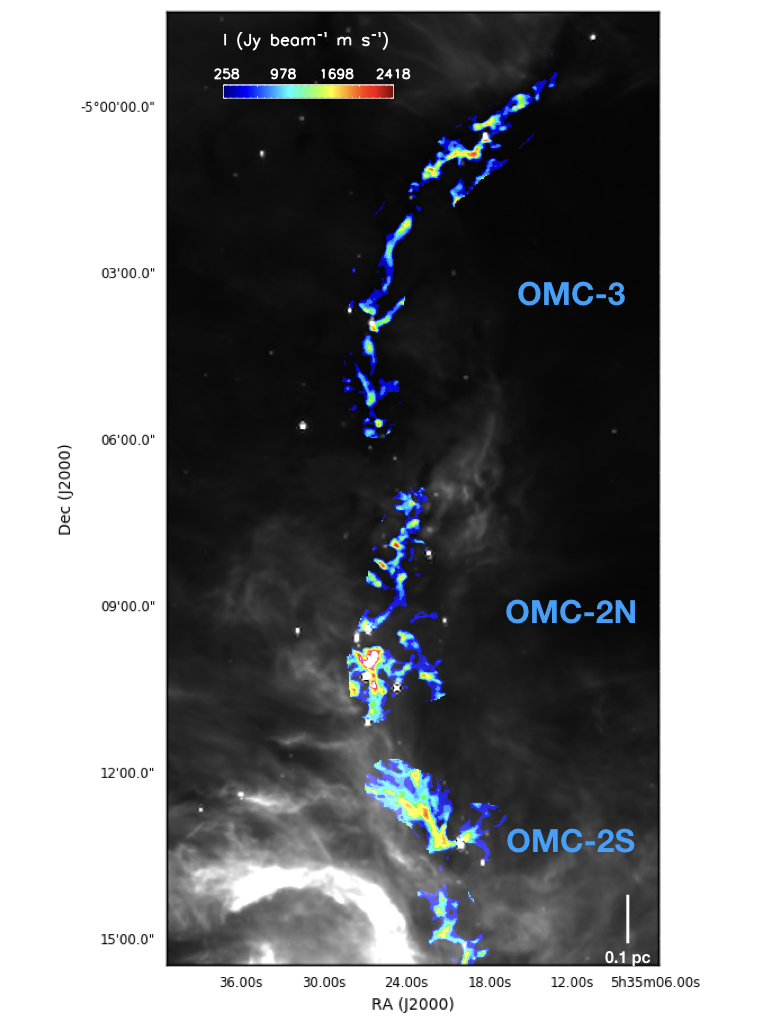}
\caption{Integrated intensity map of the N$_2$H$^+$ J=1-0 line in OMC-2 and OMC-3, overlaid on the 8$\mu$m IRAC image from the $Spitzer$ Space Telescope. The spectra were integrated between 0 and 20 ${\rm km s^{-1}}$, covering the seven hyperfine components of N$_2$H$^+$ with a 3$\sigma$ cutoff (where $\sigma$ $\sim$ 11 mJy beam$^{-1}$). Here, the spectra are from the data taking by combining the single-dish Nobeyama image into the ALMA image. The velocity resolution is 0.11 ${\rm km s^{-1}}$ and the synthesized beam is 3.46\arcsec$\times$2.20\arcsec, P.A. = 87.04$^{\circ}$. The color is scaled linearly with a range of 258 to 2418 mJy beam$^{-1}$ km s$^{-1}$ (including more than 99.5$\%$ of the data). The highest value of $\sim$3800 mJy beam$^{-1}$ km s$^{-1}$ appears in the south part of the OMC-2N region.
    }
        \label{fig:map}
\end{figure}

At a distance of 399+/-19 pc \citep{Zucker2020}, Orion Molecular Cloud (OMC) is the nearest giant molecular cloud (GMC) hosting OB clusters. With ALMA, our 3\arcsec\ resolution corresponds to 1200 au, among the highest spatial resolution ever achieved for a high mass star forming region. We estimate the Jeans scale of the dense filament in Orion as
\begin{equation}
  \lambda_J \sim \frac{0.4}{\textrm{pc}} \frac{c_s}{0.2 \textrm{ km s}^{-1}}  [\frac{\rho}{10^3\textrm{cm}^{-3}}]^{-0.5} \sim 0.005 \textrm{ pc}, 
\end{equation}
where the number density $\rho$ was estimated to be about $10^7$ $\rm cm^{-3}$ based on our ALMA 3mm continuum images (Zhu et al.\ in preparation). The proximity of Orion and the high resolution of ALMA allow us to probe dense gas in a massive star-forming GMC down to the thermal Jeans scales. The integrated intensity map of ${\rm N_2H^+}$ based on the combined ALMA and Nobeyama images is shown in Fig \ref{fig:map},which is calculated over all hyperfine components.

We fitted the spectra with a modified version of Semi-automated multi-COmponent Universal Spectral-line fitting Engine (\scouse\footnote{\scouse \ is available at \url{https://github.com/jdhenshaw/SCOUSE}})\citep{Henshaw2016}. {\sc scouse} is an IDL package to fit complex spectra line data in a robust, systematic, and efficient way, by manually fitting the spatially-averaged spectra into Gaussian components and then using them as input to the subsequent fitting on each individual spectrum.
\nthp(1-0) has hyperfine structure (HFS). We adopted the hyperfine frequencies from \citet{pagani2009} and modified {\sc scouse} by following the HFS fitting program in CLASS \citep{Forveille1989} to fit the $\rm N_2H^+$ hyperfine structure. Central velocity, intrinsic line width, total optical depth and excitation temperature are fitted.

\begin{figure}
        \includegraphics[angle = 0, trim =0cm 0cm 0cm 0cm,width=15cm]{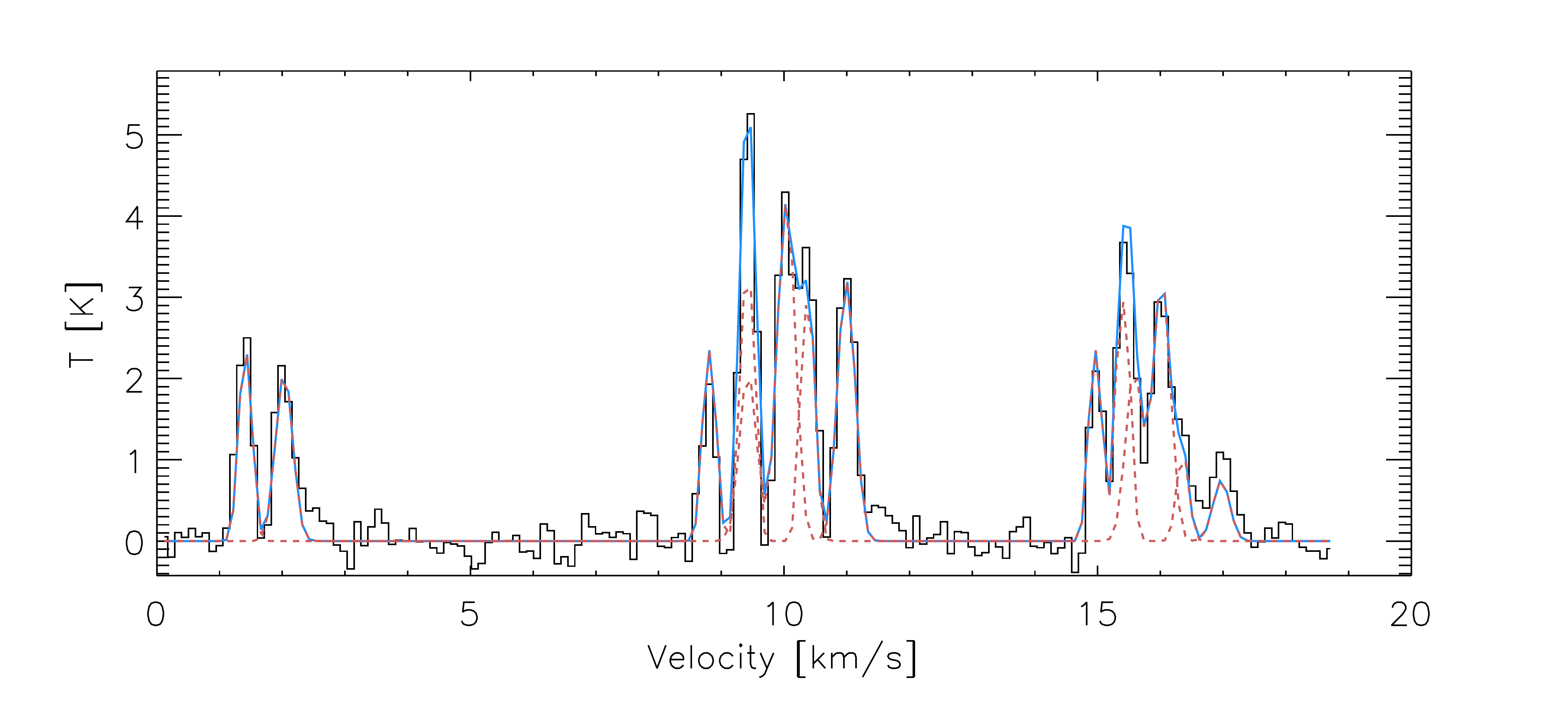}    
               \caption{One example of N$_2$H$^+$ spectrum from the position at R. A. $5^h 35^m 21^s.6$, DEC. $-5^\circ 14'38''.4$. The black line is the spectrum from the ALMA-combined  data. The red dash lines are fits to two velocity components at 1.41 and 2.03 km s$^{-1}$ (central velocity of the isolated line), respectively. The blue profile indicates the total fit to the line. The rest frequency used here is 93.1734 GHz and the LSR velocity of the center of band correspondingly is 9.4 km $\rm s^{-1}$.}
    \label{fig:example}
 \end{figure}

We applied the fitting procedure on different datasets, including the ALMA data, Nobeyama data, and the combined data. Multiple velocity components are fitted, see one example in Fig. \ref{fig:example}. Thirteen percent of the ALMA-combined data have more than one gas component, which are treated independently in the following analysis. The signal-to-noise ratio of the isolated line in ${\rm N_2H^+}$ ($I_{isolated}/\sigma_{rms}$) used as the fitting condition is 3. The mean optical depths  value of total seven hyperfine lines is 7.5 (ALMA data), 8.0 (ALMA-combined) and 2.7 (Nobeyama). We corrected the resulting line width by removing the smearing effect due to the channel width according to $\Delta\upsilon_{\rm obs}=\sqrt{\Delta\upsilon^{2}-v^{2}_{\rm chan}}$, where $\Delta\upsilon$ is the fitted line width (FWHM) and $v_{\rm chan}$ is the channel width.

\begin{figure}
        \includegraphics[angle = 0, trim =0cm 0cm 0cm 0cm,width=13cm]{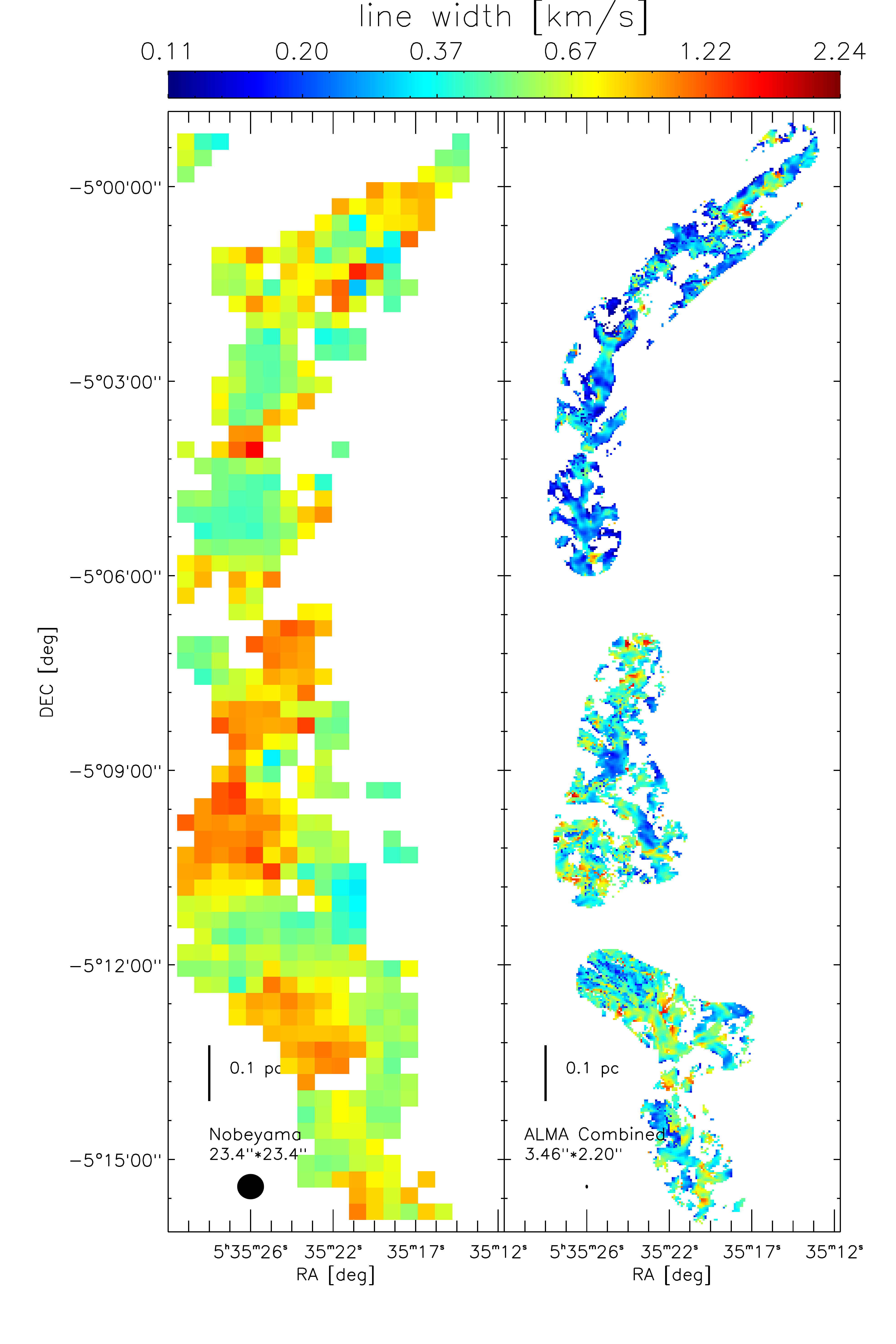}
    \caption{\textbf{Left:} line widths (FWHM) from the Nobeyama observations. \textbf{Right:} line widths from combined ALMA data. Line widths in these two panels share the same color scale. 
    }
    \label{fig:vs}
\end{figure}

\begin{figure}
        \includegraphics[angle = 0, trim =0cm 0cm 0cm 0cm,width=15cm]{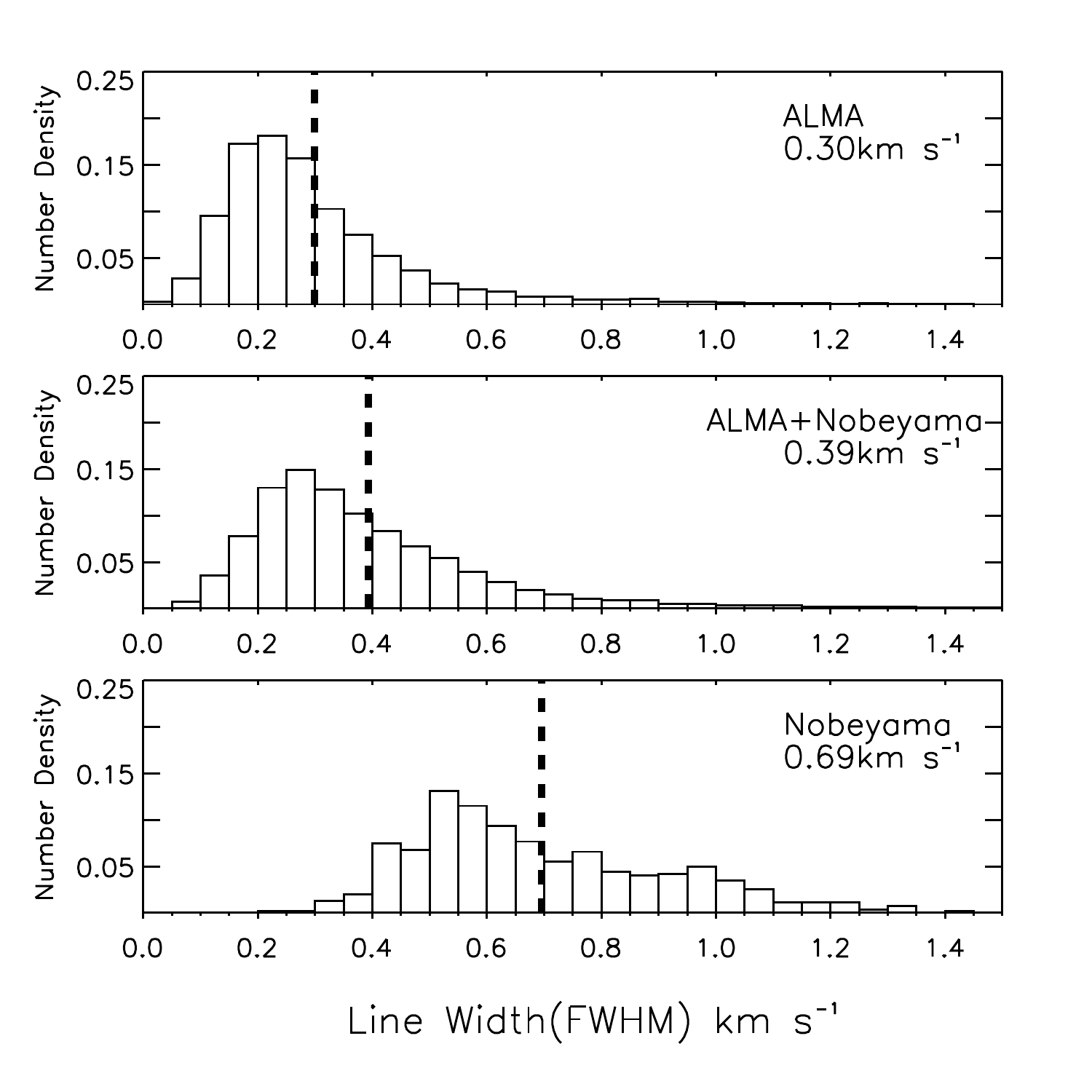}           
       \caption{ ${\rm N_2H^+}$ line width (FWHM) distributions in different datasets toward the regions OMC-2 and OMC-3. The top panel shows the number counts from the ALMA dataset (same as the ALMA row in Table 1). The middle panel shows the distribution of the combined data from Nobeyama and ALMA (same as the Combined row in Table 1). The bottom panel is data from Nobeyama observations only (same as the Nobeyama row in Table 1). The dash lines represent the mean values of each line width dataset, and the values are also labeled on each panel. The spectral line is not resolved well once the line width is smaller than 0.1 km/s (the spectral resolution).}
    \label{fig:lw}
 \end{figure}

Table \ref{tab:mean} shows the obtained mean line width (FWHM) of each region. For the entire OMC-2 and OMC-3 region, the mean line widths from the dataset of ALMA, ALMA+Nobeyama, and Nobeyama are 0.30$\pm0.01$ km s$^{-1}$, 0.39$\pm0.02$ km s$^{-1}$, and 0.69$\pm0.04$ km s$^{-1}$, respectively. The line width from ALMA is slightly smaller than that of ALMA-combined data, which suggests that the extended structures revealed by combination contribute broader line widths. The histograms of line widths obtained from each dataset are shown in Fig. \ref{fig:lw}.
The maps of line widths from Nobeyama and the ALMA-combined are presented in Fig. \ref{fig:vs}. It shows that the values of line widths from Nobeyama are mostly larger than those from the ALMA-combined data.

The non-thermal velocity dispersion $\sigma_{\rm NT}$ and the sound speed $c_s$ are also calculated using the following equations \citep{Myers1983}:
\begin{align}
(\sigma_{\rm NT})^2 = (\sigma_{\rm obs})^2 - (\sigma_{\rm T})^2; \\
\sigma_{\rm NT} = \sqrt{\frac{\Delta\upsilon^{2}_{\rm obs}}{\rm {8ln(2)}}-\frac{k_BT_{kin}}{m_{\rm obs}}};\\
c_{\rm s} = \sqrt{\frac{k_BT_{kin}}{m_{\rm mol}}}
\label{equ:disp},
\end{align}
\noindent where $\sigma_{\rm obs}$ and $\sigma_{\rm T}$ refer to the observed and the thermal dispersion, respectively. $\Delta\upsilon_{\rm obs}$ refers to the observed line width (FWHM), $k_B$ is the Boltzmann constant, $T_{kin}$ is the kinetic temperature of the gas, $m_{\rm obs}$ refers to the mass of the observed molecule (29\,a.m.u for \nthp), and $m_{\rm mol}$ refers to the ISM mean molecular weight (2.37\,a.m.u., \cite{Kauffmann2008}). 
The kinetic temperature is derived from the NH$_{3}$ inversion transitions (1,1) and (2, 2), obtained by combining JVLA and GBT observations \citep{Li2013}. 
The $T_{kin}$ used in each $\rm N_2H^+$ component is determined by the nearest position in the $\rm NH_{3}$ map.
There are certain caveats in subtracting the sound speed. The NH$_{3}$ data have a beam of 5\arcsec, while the ${\rm N_2H^+}$ data  has a beam of 3\arcsec. It is possible that the temperature may further vary within the 5\arcsec\ beam. However, within corresponding physical scale of  $\sim 0.01$ pc, the temperature variation does not produce significant line width fluctuation. Within dense gas where the gas and dust are well coupled, the gas temperatures  in dense gas at that small scale only vary by a couple of kelvins at most (see e.g.\ \cite{Li2013}). Thus if assuming the temperature at 5\arcsec\ beam is overestimate to be 14K from a real temperature of 12K, the sonic portion of the line width changes from 0.22 km s$^{-1}$ to 0.21 km s$^{-1}$. This does not affect the subsequent analysis of this paper. 

\begin{figure}
        \includegraphics[angle = 0, trim =0cm 0cm 0cm 0cm,width=15cm]{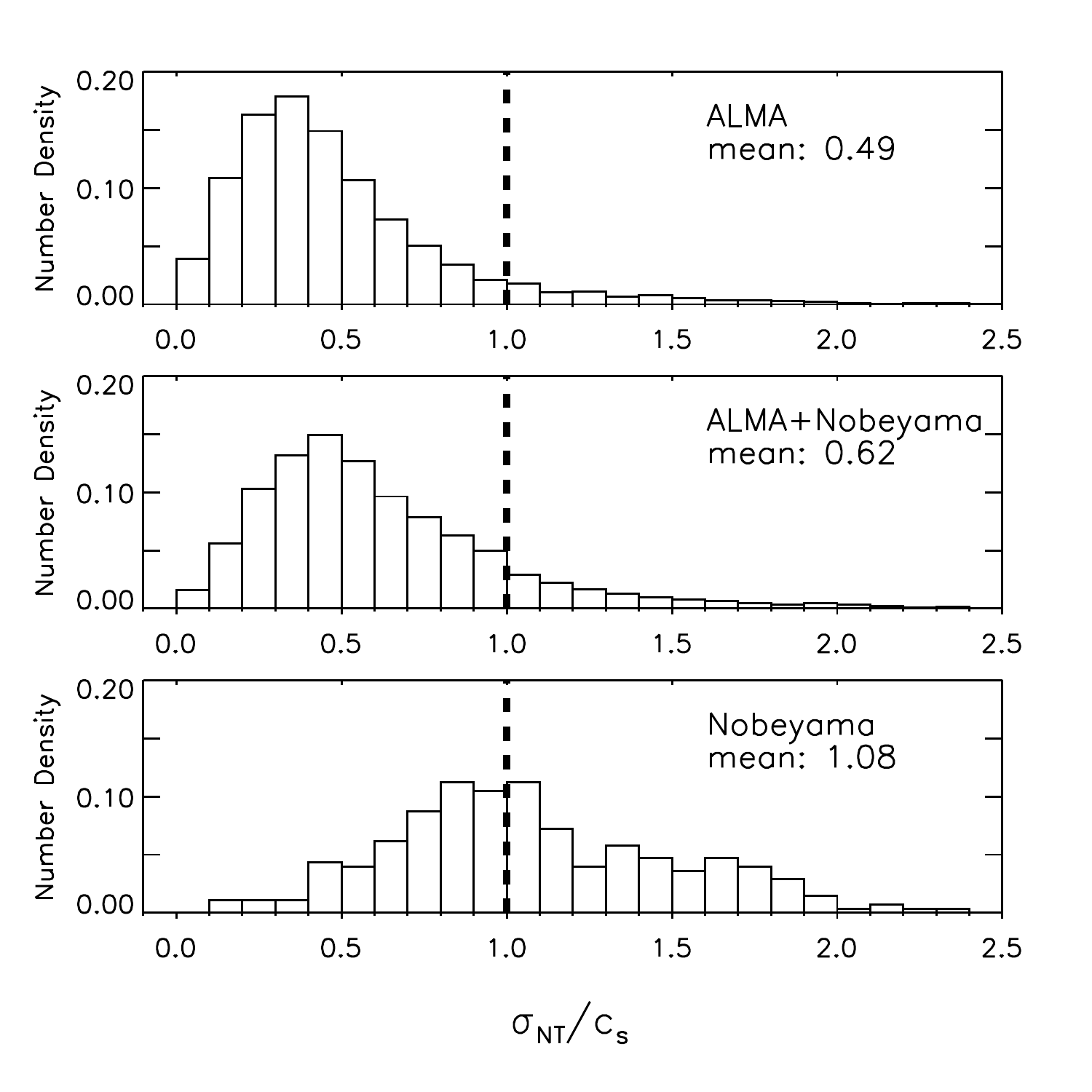}           
       \caption{Histograms of ratios between the non-thermal line dispersion and the sound speed ($\rm \sigma_{NT}/c_{s}$) for the \nthp\ 1-0 line. The mean values of $\rm \sigma_{NT}/c_{s}$ are also labeled. The vertical dashed lines refer to \ntcs = 1, which divide the data into subsonic (left of the line) and supersonic (right) regime. The sound speeds, $c_s$, are derived based on the temperature map from the VLA observations in \cite{Li2013}. 
}
    \label{fig:nth}
 \end{figure}

 A comparison of different datasets is shown in Fig. \ref{fig:nth}. For the dataset of ALMA, Combined, and Nobeyama, mean $\sigma_{\rm NT}$/$c_{s}$ values are 0.49, 0.62, and 1.08, respectively. This indicates that the line widths in the filaments from ALMA are subsonic. The line widths from the combined observations of ALMA and Nobeyama are larger but remain subsonic. In contrast, the line widths obtained from the Nobeyama data are transonic ($\sim 1$). The commonly observed supersonic or transonic line width, mostly referred to as turbulence, is a superposition of motions at different scales, which starts to be resolved by ALMA.

\begin{table}
\bc
\small
\caption[]{Mean Line Width (km s$^{-1}$) \label{tab:mean}}
\begin{center}
\begin{tabular}{c|cccc}
\hline
\hline
   & OMC-3 &  OMC-2N & OMC-2S  & OMC23$^{\rm (1)}$\\
\hline
ALMA$^{\rm a}$ & 0.27 & 0.34 & 0.29 & 0.30 \\
Combined$^{\rm b}$ & 0.30 & 0.45 & 0.45 & 0.39 \\
Nobeyama$^{\rm c}$ & 0.65 & 0.78 & 0.69 & 0.69 \\
\hline
\end{tabular}
\end{center}
\ec
\tablecomments{0.86\textwidth}{Line width here is as the full width at half maximum, FWHM.
$^{\rm (1)}\,$ OMC23 represents the entire region including OMC-2S, OMC-2N, and OMC-3. The line widths in this column show the collections from these three regions.
$^{\rm a}\,$Results from the ALMA (the 12-m main array and ACA ) observations, with a beam of $\sim 3$\arcsec.
$^{\rm b}\,$Results from the combined ALMA (the 12-m main array and ACA) and Nobeyama observations, with a  beam of $\sim 3$\arcsec.
$^{\rm c}\,$Results from the Nobeyama 45m Telescope, with a  beam of $\sim23.4$\arcsec.}
\end{table}

\begin{figure*}[ht!]
    \begin{center}
           \includegraphics[angle = 0, trim =0cm 0cm 0cm 0cm,width=15cm]{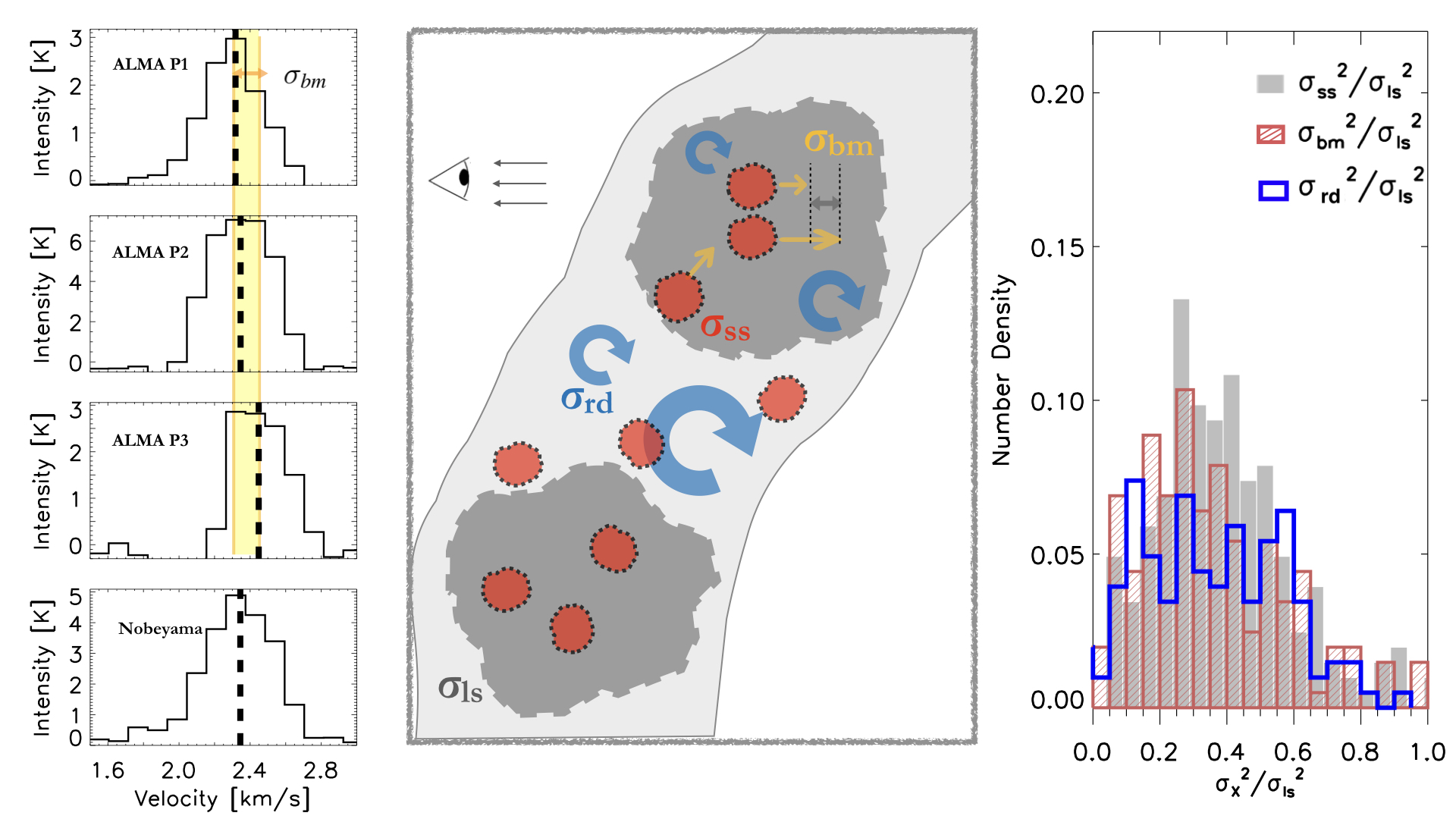}         
       \end{center}
 \caption{Illustration of resolution-dependent turbulence. \textbf{Left:} the isolated hyperfine component of the ${\rm N_2H^+}$ 1-0 line from the ALMA-combined data and Nobeyama data(bottom) at the position R. A. $5^h 35^m 26^s.4$, DEC. $-5^\circ 5'16''.8$. The ALMA spectra were sampled from within the corresponding Nobeyama beam at the position. The dashed lines are the location of centroid velocities. \textbf{Middle:} Line-dispersion components in Equation \ref{equ:energy}. The grey clumps represent structures unresolved by Nobeyama with large-scale line dispersion ($\sigma_{ls}$). The red ones represent ALMA pixels with $\sigma_{ss}$. The blue circular arrows demonstrate turbulent vortex motion in $\sigma_{rd}$. The relative motion between small ALMA clumps is shown as the yellow arrows, the dispersion among which has been labeled as $\sigma_{bm}$. \textbf{Right:} Histograms of fractions of three line dispersion components ($\sigma_{ss}^2$, $\sigma_{bm}^2$, $\sigma_{rd}^2$) to the large-scale turbulence ($\sigma_{ls}^2$). The mean radio of $\rm \sigma_{ss}^2$, $\sigma_{bm}^2$ and $\rm \sigma_{rd}^2$ is $36\%$, $56\%$, and $8\%$.  }
    \label{fig:sch}

 \end{figure*}

Further inspections of the ALMA data reveal that the line dispersion can be consistently decomposed into three components (Fig.\ \ref{fig:sch})
\begin{align}
\rm \sigma_{ls}^2=\rm \sigma_{ss}^2+\sigma_{rd}^2 +\rm \sigma_{bm}^2 \>,
\label{equ:energy}
\end{align}
where $\rm \sigma_{ls}$ and $\rm \sigma_{ss}$ refer to the line dispersion at large and small scales as measured by Nobeyama and ALMA-combined, respectively; bulk-motions $\rm \sigma_{bm}$ is the standard derivation of centroid velocities of the ALMA spectra within a single Nobeyama beam; residual  $\rm \sigma_{rd}$ represents the residual part by subtracting $\rm \sigma_{ss}$ and $\rm \sigma_{bm}$ from $\rm \sigma_{ls}$.
These components are calculated by the following equations:
\begin{align}
\rm \sigma_{ls}^2=(\rm \sigma^{nb})^2,\\
\rm \sigma_{ss}^2=\sum\limits_{i=1}^N(\rm \sigma^{alma}_{i})^2 \frac{J^{alma}_{i}}{J^{alma}},\\
\rm \sigma_{bm}^2 =\sum\limits_{i=1}^N(\rm \upsilon^{alma}_{i}-\overline{\upsilon}^{alma})^2\frac{J^{alma}_{i}}{J^{alma}}.
\label{equ:energy1}
\end{align}
$ \rm \sigma^{nb}$ is the line dispersion observed by Nobeyama, whose beam covers N pixels in ALMA observation ($\rm \sigma^{alma}_{i}$). The line dispersion at small scales, $\rm \sigma_{ss}$, is the mean line dispersion among N ALMA pixels weighted by the factor $J^{alma}_{i}/J^{alma}$. $J^{alma}_{i}$ refers to the integrated total intensity in pixel $i$ from ALMA and $J^{alma}$ is the sum of integrated intensities from N ALMA pixels within the corresponding Nobeyama area ($J^{alma}=\sum\limits_{i=1}^NJ^{alma}_{i}$).  
The bulk-motions, $\sigma_{bm}$, is described by the derivation of centroid velocities of the ALMA spectra from the mean velocity ($\overline{\upsilon}^{alma}=\sum\limits_{i=1}^N\upsilon^{alma}_{i}\frac{J^{alma}_{i}}{J^{alma}}$). $\sigma_{rd} $ is the remaining part subtracting  $\sigma_{ss}$ and  $\sigma_{bm}$ from  $\sigma_{ls}$. Averaged over the whole map, the three components all have a significant contribution to the line dispersion, specifically, with mean values of $\rm <\sigma_{ss}^2>$ at $36\%$, the bulk motion $<\sigma_{bm}^2>$ at $56\%$, and $\rm< \sigma_{rd}^2>$ at $8\%$.

Although the exact makeup of line widths varies from beam to beam (see in the right panel of Fig.\ \ref{fig:sch}), the three components are omnipresent. Checking the whole region, where the quality of combination may change from a total flux recovery at strong emission points to lower flux recovery ($\sim 70\%$) at diffuse regions, the residual part, $\rm \sigma_{rd}^2 $ is still present. This may, to some extent, show that the existence of the residual part is not because of the imperfection of data reduction.
We further postulate that, when the ALMA beam is resolved with finer resolution, $\sigma_{ss}$ can be broken into three components, i.e., the non-thermal line dispersion currently obtained at 1200 au scales (albeit subsonic) can still be further decomposed into relative motions and seemingly `intrinsic' line width. Such a testable hypothesis could help reveal a fundamental quality of the dense gas motion in a massive star-forming cloud, namely, the turbulence (as seen by line width) seems to be self-similar, but resolution-dependent to observers.

\section{Discussion}
Massive stars can form in supersonically turbulent gas through turbulent core accretion \citep{McKee2002,McKee2003}. \citet{Myers2014} also simulated star formation in turbulent, magnetized dense clumps with radiative and outflow feedback. In their simulations, the turbulence creates a network of overdense filamentary structures, where  stars are born.

Recent high-resolution observations, however, consistently find narrow line width.
\citet{Sokolov2018} found 1/3 of pixels in their \nht\ images with subsonic non-thermal line dispersion. Compared to this work, that data set has one order of magnitude coarser spatial resolution at about 0.07 pc. This corresponds to the transition scale found in this work. Furthermore, the line width analysis in \citet{Sokolov2018}'s was largely based on hyperfine fitting of \nht (1,1), which has unresolved components. Previous work \citep{Stutzki1985,Li2013} have shown that the fitted opacity and line width can be coupled and affected by blending.
\citet{Hacar2018} did observe the N$_2$H$^+$ (1-0) line toward OMC-1 and OMC-2 with ALMA and also found the non-thermal velocity dispersion similar to or smaller than that of the thermal motion. They combined ALMA and IRAM 30m data and found a mean \ntcs\ of 0.81 ($>$$75\%$ of the positions are subsonic).
Our observation is consistent with this trend, but yet have substantially better fidelity at finding subsonic turbulence.
In direct comparison with \citet{Hacar2018}, our work utilized combined GBT+VLA NH$_3$ data images (5\arcsec) vs.\ GBT-only ones with a 32\arcsec\ beam in \citet{Hacar2018}, which could result in overestimation of sound speed. 
Our ALMA data are also more than 2 times deeper ($\sim$ 11 mJy/beam) than those in \citet{Hacar2018} (25 mJy/beam) and thus provide better dynamic range. 
   
\citet{Megeath2016} presented more than 5000 YSOs in both Orion A and Orion B. Based on the same data sets, we have discussed a subset of protostars in OMC2/3 in \citet{Li2013}. These protostars are predominantly low mass objects, although the thermal structure of the whole Orion-A cloud is purportedly determined by the UV field from the OB cluster \citep{Stacey1993,Li2013}. The influence of the Trapezium cluster is manifested in elevated cloud surface temperatures, which appear to be decreasing with distance to the cluster. Even under the influence of an OB cluster, the dense gas in Orion still transits to subsonic non-thermal line widths, when properly resolved.
The unequivocally detected subsonic non-thermal line dispersion here shows that Orion, an OB cluster hosting region, has similar dependence of line dispersion on spatial scales to that in low mass star forming regions \citep{Goldsmith1987,Caselli2002}.

Previous studies, mostly in low mass star forming regions, seem to be able to identify 
spatially coherent features, where the transition of subsonic line width happen. \cite{Pineda2010} presented a transition to the velocity dispersion between a subsonic dense core and its surrounding gas by observing $\rm NH_3$ towards B5 region in Perseus with GBT. They found a factor of 2 more turbulent gas in the surrounding environment than in the dense core. Further high resolution observations found an subsonic filament and dense gas condensations embedded within this region \citep{Pineda2010} and suggested that filament fragmentation can lead to multiple stellar systems by  \cite{Pineda2011}. \cite{Hacar2013} found transonic filament bundles in the L1495/B213 Taurus region.
No such transition feature, either in filaments or core envelopes, can be  clearly seen in OMC2 and OMC3.
More detailed analyses are being carried out.

  \section{Conclusions}
 
 In this paper, we report ALMA observations of the ${\rm N_2H^+}$ J=(1-0) transition toward OMC2 and OMC3, the relatively quiescent
region in the Orion molecular cloud, which is the nearest giant molecular cloud with OB clusters. .
 Our main results are as  follows:
 
 1 We obtained  high-quality ALMA images of dense gas (${\rm N_2H^+}$) in Orion Molecular Cloud 2 and 3  with an RMS of 11 mJy per $\sim 3$\arcsec\ beam ($\sim$0.006 pc) per 0.11 km s$^{-1}$ channel, which are more sensitive than previous works of comparable spatial resolution (cf.\ \citet{Hacar2018,Ren2014}). We were able to probe density peaks as well as  extended structures down to the thermal Jeans scale (0.005 pc), 
with unprecedented spatial dynamic range.
 
 2 We analyzed the non-thermal component of the line width, based on a combined analysis of both ALMA+Nobeyama ${\rm N_2H^+}$ data and GBT+VLA ${\rm NH_3}$ images with comparable resolution. There exists a clear transition from transonic non-thermal line dispersion ($\sigma_{\rm NT}$/$c_{s}$=1.08) to subsonic line dispersion ($\sigma_{\rm NT}$/$c_{s}$=0.62) between the scales of $\sim 0.05$ pc and $\sim 0.006$ pc.

 3 The line dispersion of each single-dish beam (Nobeyama) can be decomposed into three components, namely, small-scale line dispersion $\sigma_{ss}$ (based on ALMA-combined) of $36 \%$, bulk motion $\sigma_{bm}$ (based on the dispersion among the centroid velocities of each ALMA-combined beam) of $56 \%$, and the residual part $\sigma_{rd}$ of 8 \% being the persistent difference between the sum of the two aforementioned components and the full line width from the single-dish beam.
 
4 The transition from supersonic to subsonic line width (`turbulence') suggests that supersonic turbulence may be not necessary in massive star formation at certain scales.

\begin{acknowledgements}
This work is supported by the National Natural Science Foundation of China Grant No. 11988101, No. 11725313 and No. 11629302, the CAS International Partnership Program No.114A11KYSB20160008, This work was carried out in part at the Harvard-Smithsonian Center for Astrophysics .
\end{acknowledgements}

\bibliographystyle{aa}
\bibliography{myrefs}

\end{document}